\documentclass[conference]{IEEEtran}
\ifCLASSINFOpdf
\else
\fi
\usepackage[utf8]{inputenc}
\usepackage{xcolor}
\usepackage{hyperref}
\usepackage{amsmath, amsthm, amssymb}
\usepackage{url}

\usepackage{graphicx}

\usepackage[margin=1in]{geometry}

\definecolor{darkgreen}{rgb}{0.0, 0.5, 0.0}
\definecolor{denim}{rgb}{0.08, 0.38, 0.74}

\usepackage{pifont}

\usepackage{lipsum}

\begin{document}
%
\title{Differentially Private Health Tokens for Estimating COVID-19 Risk}



%

\author{\IEEEauthorblockN{David Butler$^*$\IEEEauthorrefmark{2},
Chris Hicks$^*$\IEEEauthorrefmark{2},
James Bell\IEEEauthorrefmark{2}, 
Carsten Maple\IEEEauthorrefmark{2}\IEEEauthorrefmark{3} and
Jon Crowcroft\IEEEauthorrefmark{2}\IEEEauthorrefmark{4}}
\IEEEauthorblockA{Email: \{dbutler, chicks, jbell, cmaple, jcrowcroft\}@turing.ac.uk}
\IEEEauthorblockA{\IEEEauthorrefmark{2}Alan Turing Institute, \IEEEauthorrefmark{3}Univerity of Warwick, \IEEEauthorrefmark{4}University of Cambridge}
}



\maketitle

\begin{abstract}
In the fight against Covid-19, many governments and businesses are in the process of evaluating, trialling and even implementing so-called immunity passports. Also known as antibody or health certificates, there is a clear demand for any technology that could allow people to return to work and other crowded places without placing others at risk. One of the major criticisms of such systems is that they could be misused to unfairly discriminate against those without immunity, allowing the formation of an `immuno-privileged' class of people. In this work we are motivated to explore an alternative technical solution that is non-discriminatory by design. In particular we propose health tokens --- randomised health certificates which, using methods from differential privacy, allow individual test results to be randomised whilst still allowing useful aggregate risk estimates to be calculated. We show that health tokens could mitigate immunity-based discrimination whilst still presenting a viable mechanism for estimating the collective transmission risk posed by small groups of users. We evaluate the viability of our approach in the context of \textit{identity-free} and \textit{identity-binding} use cases and then consider a number of possible attacks. Our experimental results show that for groups of size 500 or more, the error associated with our method can be as low as 0.03 on average and thus the aggregated results can be useful in a number of \textit{identity-free} contexts. Finally, we present the results of our open-source prototype which demonstrates the practicality of our solution.

\end{abstract}


%

\section{Introduction}

The discussion surrounding immunity, antibodies and transmission risk have become synonymous with the fight against COVID-19. One technique that has been proposed is to certify each individuals' immunity status in a so-called immunity-passport. Because immunity to COVID-19 is not well understood, and the correlates to protection have not been fully identified, it has been proposed that these certificates (which are currently based on antibody testing) should be called antibody certificates \cite{DBLP:journals/corr/abs-2005-11833}. 

A number of antibody certificate solutions have already been proposed \cite{certus_blockchain_solution, estonia_passport, immupass,  DBLP:journals/corr/abs-2004-07376, DBLP:journals/corr/abs-2005-11833} and there has been significant discussion of the technical, social and ethical implications of the technology \cite{lovelace_report, WHO_immunit_passports, Kofler_Baylis_2020}. One of the main concerns relating to antibody certificates is that they could be misused to unfairly discriminate against people with respect to their immunity status. It has been suggested \cite{DBLP:journals/corr/abs-2005-11833} that antibody certificates should not restrict peoples access to services, or freedom of travel, but could instead be useful as a tool for measuring aggregate transmission risk levels. To this end we approach this work with two assumptions in mind:

\begin{enumerate}[\topsep=1ex]
	\item Immunity status is an attribute that should not be a basis for widespread, arbitrary discrimination. 
	\item It is useful to know the ratio (or count) of \emph{immune} vs \emph{non-immune} people that have accessed a service over a period of time --- for example customers in a shop over a number of hours.
\end{enumerate}

Our first assumption is motivated by concerns that immunity-based discrimination could threaten freedom, fairness and public health \cite{Kofler_Baylis_2020}. Particularly in the absence of a widely-available and effective vaccine: access to shops, transportation, religious establishments and medical services should perhaps not depend on whether wealth or luck has favoured an individual with immunity. We are motivated by these concerns to develop a system that technically mitigates immunity-based discrimination, using differential privacy, and to explore the limitations and practicality of our approach. Our second assumption is motivated by the idea that the presence and proximity of immune individuals may offer a proxy for estimating the risk of infection to those who are still susceptible \cite{10.1093/cid/cir007}. 

Within the framework of these two assumptions we develop the idea of a \emph{health token}: a randomised health certificate that allows the collective transmission risk\footnote{We assume that a measure of transmission risk exists and that it can be accurately determined.} posed by groups of users to be estimated whilst also protecting individuals from immunity-based discrimination. Unlike previous work on antibody certificates, randomised health tokens are less likely to be used to restrict access to services or transportation due to the uncertainty introduced in their true value. Health tokens provide each user with a transmission risk that is plausibly deniable. Consequently, there is minimal incentive for user fraud and so the need to bind each user to their token is greatly diminished. Our health tokens comprise only a digitally signed, differentially private transmission risk status that allows aggregate transmission risk statistics to be computed.

In our proposed system, the transmission risk of each user is first evaluated by a healthcare provider who then cryptographically signs and issues the user a health token. Of course, our system is independent of how this infection risk is calculated and can support any number of different risk profiles, for example a scale of $1$ to $k$. 

Like many technology-based approaches to mitigating COVID-19, the effectiveness of health tokens is dependant on the number of people willing to participate in the system voluntarily. We envisage that, like digital contact tracing apps, citizens will be motivated to use health tokens by their will to bring social distancing measures to an end and to help in the fight against the disease. 

In addition, our proposal may encourage users who would be reluctant to engage with a regular immunity-passport system that binds them directly to a certificate. Indeed our health tokens do not reveal any personal information, such as a name or photograph, which could either be used to trace a user across multiple contexts or be otherwise mishandled. Moreover the non-binding nature of our system means there is little risk of feature creep.

In Section \ref{sec:protocol} we detail our proposed system and in Section \ref{sec:use cases} we consider two types of use case: \textit{identity-free} cases such as shops and \textit{identity-binding} cases such as aeroplane flight booking. We consider and analyse some possible attacks on our system in Section \ref{sec:attacks and analysis}. Finally in Section \ref{sec:limitations} we discuss the practicalities and limitations of our approach and in Section \ref{sec:conclusion} we review the related work and conclude.


\section{Non Discriminatory Token System} \label{sec:protocol}

\begin{figure}[h]
	\centering
	\includegraphics[scale=0.18]{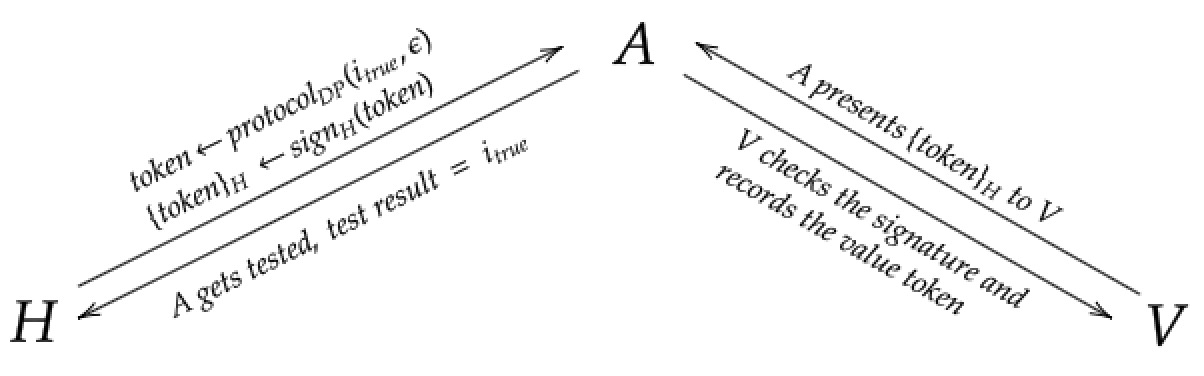}\vspace{-0ex}
	\caption{The high-level details of our health token system.}
	\label{fig:system}\vspace{-2ex}
\end{figure}

Our system consists of three phases: issue, check and aggregate. In the issue phase a healthcare provider assesses the transmission risk of a user and then issues them with a cryptographically signed health token. In the check phase the service provider records the value of the users' health token. Finally, the aggregate phase allows the service  provider to reconstruct the overall transmission risk associated with the `checked' tokens. The high-level details of our health token system are depicted in Figure \ref{fig:system}.

\subsection{The Health Token System}

Here we present the three phases of our system in turn. First we denote the three stakeholders as follows: the healthcare provider Henry (H), the user Alice (A) and the service provider Verity (V, alliterative for verifier). Let $\mathit{protocol}_{\mathit{DP}}$ denote the differential privacy protocol that we use to compute our health tokens, then the three phases of our system are as follows:
\vspace{1ex}

\noindent \textbf{Issue}
\begin{enumerate}
	\item H assesses A's transmission risk indicator to be $i_{\mathit{true}} \in \{0, \dots, k-1\}$. 
	\item H computes $\mathit{token} \leftarrow  protocol_{DP}(i_{\mathit{true}}, \epsilon)$ and produces $\{\mathit{token}\}_{\mathit{sgn_H}}$ for A.
	\item A receives $i_\mathit{true}$ and $\{\mathit{token}\}_{\mathit{sgn_H}}$.
\end{enumerate}

\noindent \textbf{Check}
\begin{enumerate}
	\item A presents $\{\mathit{token}\}_{\mathit{sgn_H}}$ to V before accessing a service.
	\item V verifies the signature and records the value $\mathit{token}$. 
\end{enumerate}

\noindent \textbf{Aggregate}
\begin{enumerate}
	\item V finds the frequency, $f$, of each element of $\{0,\dots,k-1\}$ amongst the $\mathit{token}$s it has received.
	\item V debiases the frequency estimates as per Equation \ref{eq:debias} to get an unbiased estimate of the the sum of the health tokens it has collected.
\end{enumerate}

Each token is cryptographically signed by H. In particular we denote by $\{\mathit{token}\}_{\mathit{sgn_H}}$ the user token $\mathit{token}$, signed with H's private signing key ${\mathit{sgn_H}}$. The bitstring representation of each signature will be unique (except for a negligible probability) for each token thus can be used as an identifier for each health token. We denote this identifier by $\mathit{TID}$ --- token identification number. Note that regardless of the user token value, A learns their true transmission risk.

\subsection{The Differential Privacy Protocol} \label{sec:dp_protocol}

We use the differential privacy technique of randomised response \cite{randomised_response} to produce a token based on the transmission risk of a user as evaluated by a healthcare provider. Informally, the protocol which we denote $\mathit{protocol}_{\mathit{DP}}$ is as follows.

The input is the users' true risk status $i_{\mathit{true}} \in \{0, \dots, k-1\}$, which can be one of $k$ options, and $\epsilon$. Here, $\epsilon$ is a measure of the trade-off between privacy and accuracy; a low value of epsilon provides high  privacy, but low utility (as the underlying distribution is more random) and vice versa. The protocol outputs $i_{\mathit{true}}$ with probability $\frac{e^{\epsilon}-1}{e^{\epsilon}+k-1}$, else we select a uniformly random response from $\{0, \dots, k-1\}$. Clearly it is possible to select a response at random and to still end up providing the true value. 

An unbiased estimate $\hat{f}$ of the frequency $f$ of a given option is calculated by putting the frequency it is returned as a response, $\tilde{f}$, into the following formula:
\begin{equation}\label{eq:debias}
\hat{f}=\frac{e^{\epsilon}+k-1}{e^{\epsilon}-1}(\tilde{f}-1/k).
\end{equation}

Error in re-aggregating the data is, of course, introduced with such a method. We visualise this error in our experiments in Section \ref{sec:conclusion}. 

\section{Use Cases} \label{sec:use cases}

Here we first give an example use case of our system, specifically we show how a shop could implement a risk analysis using health tokens. We then discuss the situations we do not think health tokens are appropriate; we conclude that health tokens are only suitable for \emph{identity free} use cases.

\subsubsection*{Identity-Free Use Case – Allowing users into a shop}

We consider a shop that wants to monitor the level of transmission risk inside at any given time. We construct a rudimentary risk model that demonstrates how our system could add benefit.  

Let $R$ be the natural number that represents the upper bound on acceptable risk in a shop at any one time and let $T$ be the set of health tokens associated to people currently in the shop. Then $$r = \sum_{t \in T} t$$ is a measure of the total transmission risk in the shop\footnote{Note this has an error associated with it due to the randomness of the health tokens introduced in the Issue phase.}. We let $t_n$ and $t_e$ denote the token value of the next customer (in the queue to enter) and the exiting customer (the next customer to exit the shop). The next customer is let if and only if $r - i_e + i_n \le R$.

Moreover it may be beneficial to aggregate the cumulative level of risk the shop has been exposed to over the course of a day or week --- this may help guide when cleaning is most required, for example.

We reiterate that the risk model we have presented here is primitive. The method however could be abstracted to any risk model that is based on the customers token value. For example, it could take into account the size of the shop, the current staffing levels, or even the current spatial distribution of customers in the shop. 

The example above suggests the utility of health tokens in adding a net positive benefit to society. Comparable uses could be monitoring the risk of crowds at sports stadiums and classes in schools.

\subsubsection*{Limits on Use Cases - Identity-Binding}

We identify that not all situations are suitable for health tokens. Consider the example of filling seats on an aeroplane based on transmission risk. The objective function is ill-defined however the airline would likely want to limit risk while also filling as many seats as possible.

One can imagine a process where at the time of check-in, a passenger must also submit their health token value and the airline computes an aggregated risk value in the same way as a shop does. At this point an issue arises: the users' health token is now bound to their identity through their passport. We highlight two immediate negative consequences of this: (1) the binding can be remembered and could be used to unfairly impact users with unfavourable health token values such as when purchasing insurance or booking future flights and (2) airlines are incentivised to allow only low risk passengers on-board (to maximise flight capacity) thus reintroducing discrimination.

The problem here is caused by the \emph{identity-binding} of the health token. While all situations are different, and thus in some \emph{identity-binding} situations there may be possible solutions we conclude that health tokens are only suitable for \textit{identity-free} use cases.


\section{Attacks and Analysis} \label{sec:attacks and analysis}

In this section we first define numerous attack vectors against our system and then analyse their effect and possible mitigations.

\subsection{Attack Vectors}
All distributed systems are susceptible to Sybil attacks \cite{sybil} in which an adversary impersonates another user or otherwise exploits the lack of binding between users and credentials. After Sybil attacks, we also consider adversaries who wish to learn about a user based on observing their behaviour or demanding to see their health token. We define a compromised health token to be one that is used without user consent.

\paragraph{Sybil-type A}

A Sybil-type A adversary attempts to disrupt a service provider by repeatedly presenting a set of compromised tokens to a service provider. For example, if the adversary presents tokens that indicate a high level of risk, the service provider may be forced to close or act more cautiously. 

\paragraph{Sybil-type B}

A Sybil-type B adversary builds on the type A adversary. Here the adversary attempts to use a set of compromised tokens to attack multiple service providers. For example, an adversary could corrupt real user's phones with malware that overwrites the original certificate to the one of their choosing. The adversary's goal here could be to change government policy by artificially altering the aggregated transmission risk at a national scale. 

\paragraph{Queue Observer}

This adversary is interested in inferring a persons true transmission risk status by monitoring when they are allowed (and when they are refused) entry to services. We assume this adversary may be interested in monitoring a user over a long period of time.

\paragraph{Malicious Service Provider}

This adversary is interested in learning a user's true transmission risk status and potentially discriminating against them based on its value. While it is likely such an adversary will be a service provider, our definition also however includes unauthorised or coercive peers.


\subsection{Analysis of Attacks}

When analysing the Sybil-type attacks we assume the adversary does not have access to a large number of valid health tokens --- that is we assume they only have access to a small number of compromised certificates which they can use. On one hand this is reasonable as a valid certificate is signed by H and thus cannot be forged. On the other hand since our health tokens are not binding, a malicious user can easily use a token that was issued to someone else. Indeed, a malicious user could post their health token online for others to use should they choose. Under this informal assumption, which it is important to debate beyond this work, we show how Sybil-type attacks can be mitigated.

\paragraph{Sybil-type A}

As our health tokens are non-binding, an adversary may attack a service provider by repeatedly presenting a compromised health token of their choosing. A determined adversary may even convince many users, unsuspecting or otherwise, to present a singular compromised token that has been distributed online for this purpose.

To prevent this attack a service provider can simply count the number of times each $\mathit{TID}$ is seen in the period of a day or week. In this way a limit can be imposed on the number of times each certificate can be used over a given period of time. For an adversary to effectively use such an attack despite this mitigation they would need access to a large number of signed tokens with their desired value. In practice they would also need to convince or trick many individuals into using them.

\paragraph{Sybil-type B}

By rate-limiting the number of uses of each token with respect each service provider, the adversary is limited to attempting to present each compromised token they hold the maximum number of times and to as many different service providers as possible. This adversary might try to alter the overall risk aggregates across a country, or on a smaller scale might try all the shops in a particular shopping mall or tube station. 

To prevent this more widespread attack we need to monitor the usage of a certificate across multiple service providers. There are at least two methods of achieving this: (1) a central body requests and aggregates the usage count of each health token across multiple service providers and (2) the service providers communicate directly with each other to learn which certificates are being used the most --- here we envisage a multi-party computation protocol might be used to do so in a privacy-preserving way.

While it is often desirable to avoid using a central authority where possible, it is likely that method (2) will both involve too large a communication overhead and place too much burden on service providers. Therefore in this work we only consider solutions using method (1). Naor et al.~\cite{DBLP:conf/ccs/NaorPR19} present a secure method to find `heavy hitting passwords' (i.e. commonly used passwords) using a central authority. This solution can also be applied here to allow a central authority to learn, in a privacy-preserving way, if there are any health tokens with an unreasonably large usage across multiple service providers. We outline the details of this approach for this application in Appendix \ref{appendix:naor protocol}.

\paragraph{Queue Observer}

No adversary can be confident of knowing the users true transmission risk status because the randomised response that is applied to each health token provides plausible deniability. However, an adversary who tracks a user across multiple service providers could learn (with a degree of uncertainty) a users token value. 

To stop an adversary that observes multiple queues, service providers or policy makers could introduce batch admission to services. With this proviso, any information an adversary can learn by observing a user is masked by the other users in the batch.

\paragraph{Malicious Service Provider}

By definition, health tokens are plausibly deniable and protect against immunity-based discrimination.

\section{Limitations and Implementation} \label{sec:limitations}

Here we discuss the limitations of our proposed approach and then demonstrate the practicality of our solution by presenting the results of our open-source implementation which can be found at \cite{implementation}. 

\subsection{Limitations}

\paragraph{Error in aggregation}

Our system perturbs a users true transmission risk with some randomness to produce a health token value. 
It is essential that health tokens are only used when considering a large sample of users and not for individual cases. For example, determining if a specific user is allowed into a care home is probably best left to a system with strong user-certificate binding such as \cite{DBLP:journals/corr/abs-2005-11833}.

We illustrate the error associated with our method by calculating the average error in computing the expected value of a health token ($E[X]$) for a set number of users. First we let $\epsilon = \; \mathrel{log}3$, which corresponds to outputting the correct risk status in the first step of the  underlying differential privacy protocol (as described in Section  \ref{sec:dp_protocol}) with probability $\frac{1}{2}$, and show how the error changes as we increase $k$. Second  we keep $k = 2$ constant and vary $\epsilon$ to allow for $\frac{1}{4}, \frac{1}{2}$ and $\frac{3}{4}$ chance of outputting the truth after the first (bias) coin flip  in the underlying differential privacy protocol. We iterate the experiments 100 times for each number of users from 1 to 500. We see that, as expected, increasing the number of options for risk status value (increasing $k$) results in a larger error, as does increasing $\epsilon$.

\begin{figure}[h]
\vspace{-0ex}
	\centering 
	\includegraphics[scale=0.35]{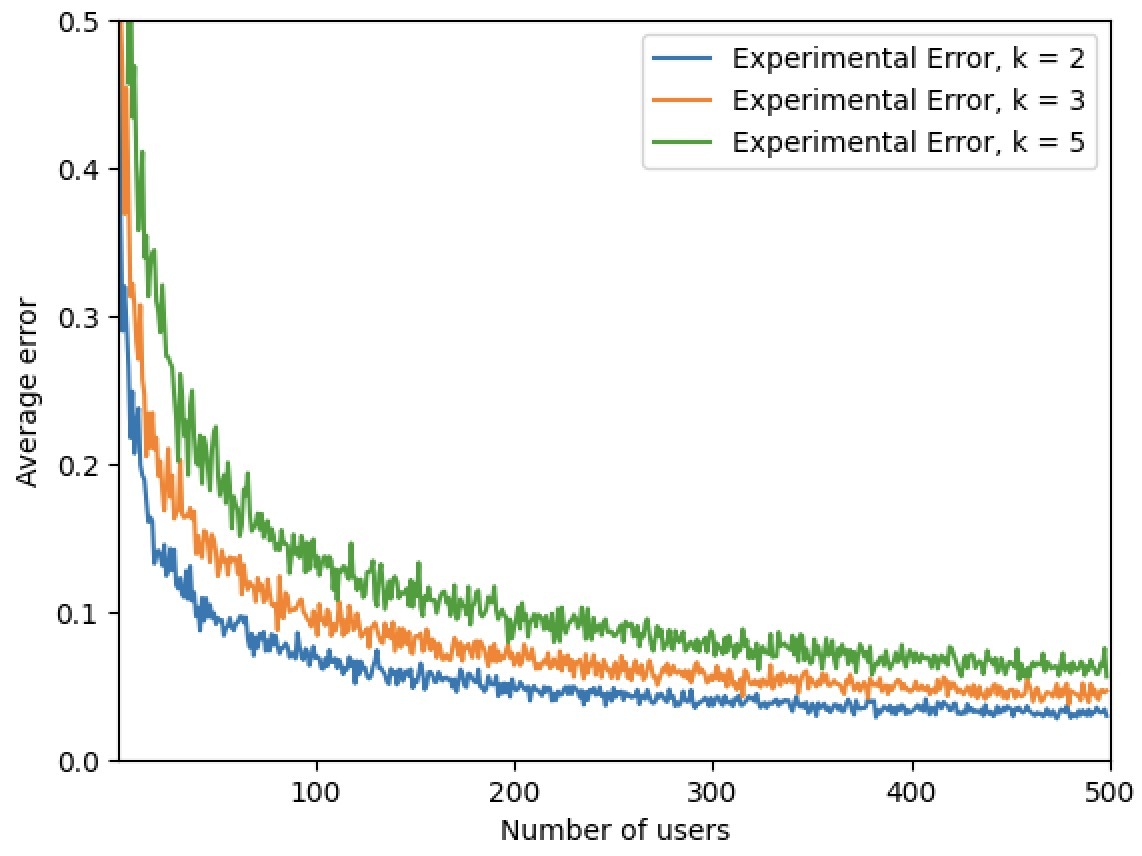}\vspace{-1ex}
	\caption{The average error introduced by our system for a given number of users. We let $\epsilon = \; \mathrel{log}(3)$ and plot the error for $k = 2, k = 3$ and $k = 4$.}
	\label{fig:average_error2}
\end{figure}\vspace{-1ex}
\begin{figure}[h]
	\centering 
	\includegraphics[scale=0.48]{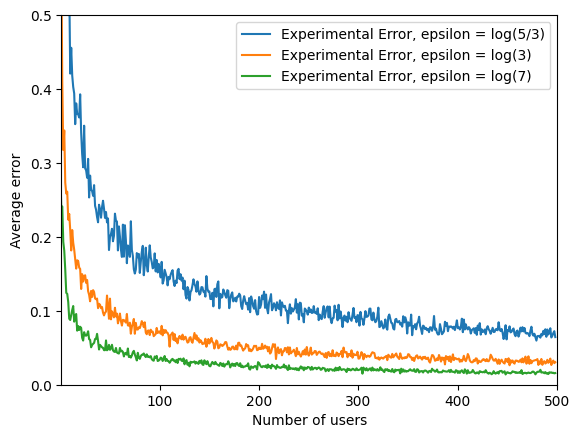}\vspace{-1ex}
	\caption{The average error introduced by our system for a given number of users. We let $k = 2$ and plot the error for $\epsilon = \; \mathrel{log}(\frac{5}{3}), \epsilon = \; \mathrel{log}(3), \epsilon = \; \mathrel{log}(7)$ and $k = 2$.}
\label{fig:average_error1}
\end{figure}

\paragraph{No binding}

As the tokens are not binding, their misuse is harder to mitigate or track. As the Sybil-type attacks show, it is possible that compromised certificates could be used maliciously by adversaries. It is also possible that a user will be incentivised to be tested repeatedly until they get issued a more \emph{desirable} health token value; In other words a user with an \emph{undesirable} risk status may be incentivised by our system to test again (or to bribe their healthcare provider). 


\paragraph{User behaviour}

We cannot fully anticipate the impact of our system on user behaviour. For example, users assigned high risk values may choose not to present their health token to service providers. It may be difficult to anticipate, measure and correct for this bias. 

\paragraph{National aggregation and reporting}
In our system, service providers keep track of their local aggregate risk value and how many users have been in their environment during a specific period. We have not considered in this work how  multiple service providers can aggregate their results and contribute to a wider geographical or national risk model. This is challenging because a service provider could be incentivised to report misinformation, particularly if restrictions on trade could be imposed as a result of reporting a high risk measurement. While it could be difficult to verify each service providers reporting, it would be possible to spot providers that continually misreport relative to their peers.

\subsection{Implementation}

\begin{figure}[h]
	\centering 
	\includegraphics[scale=0.21]{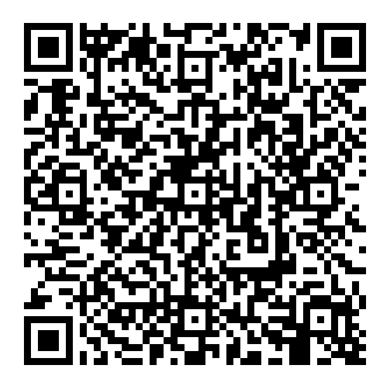}\vspace{-0ex}
	\caption{Example token generated by our prototype.}
	\label{fig:qr_code}
\end{figure}

We implement our system in Python using using QR codes as a means of presenting and verifying a signed health token. As shown in Figure \ref{fig:qr_code}, our prototype allows for the generation and verification of a health token signed using 512-bit ECDSA \cite{FIPS186}.

\section{Related Work and Conclusion} \label{sec:conclusion}

\subsection{Related Work}
To the best of our knowledge this is the first antibody certificate system that, by design, mitigates against immunity-based discrimination. Other proposed systems are dependant on policy and legal measures to achieve the level of non discrimination achieved by our proposal. 

In \cite{DBLP:journals/corr/abs-2005-11833}, the secure antibody certificate system (SecureABC) was introduced that uses a standard public-key signature scheme to ensure the binding and authenticity of certificates. Likewise in \cite{Eisenstadt} a system composed of W3C-standard `verifiable credentials' \cite{sporny2019verifiable}, the `Solid' platform for decentralised social applications \cite{DBLP:conf/www/MansourSHZCGAB16} and a federated consortium blockchain was developed. In both these works however the user is strongly bound to their certificate. In fact it is implied that certificates are only issued to those who \emph{pass} a test, thus not holding a certificate is synonymous with not having immunity and therefore not being immuno-privileged. 

Therefore we suggest these identity-bound systems, along with similar industry led proposals \cite{certus_blockchain_solution, immupass} are more applicable to situations where immunity-based discrimination can be justified. For example for workers in a care home and the identity-binding use cases we considered in Section \ref{sec:use cases}. 


\paragraph{Conclusion}

In this work we were motivated to develop a tool for measuring the viral transmission risk of groups of users which, by design, mitigates immunity-based discrimination. While there are a number of limitations to our proposed health tokens, we have shown that in identity-free situations our approach has the potential to have a net-positive benefit on society.



%
\bibliographystyle{plain}
\bibliography{main}

\appendix

\section{Preventing Sybil-type B attacks} \label{appendix:naor protocol}

Naor et al. \cite{DBLP:conf/ccs/NaorPR19} present a solution to the problem of identifying commonly used passwords – we propose their method for out situation. Similarly to identifying password outliers with respect to their frequency of use, we require a method to identify health tokens that are used so frequently that adversarial behaviour is indicated. Here we present the original solution of Naor et al. for completeness. A central authority can securely aggregate the count of health token TIDs as follows:

\begin{enumerate}
    \item For each health token, V hashes the TID, $v_j = H(\mathit{TID_j})$ --- assume this is an $l$-bit string, in practice we suggest $l = 32$.
    \item The server sends V a uniformly distributed $l$-bit string $r_j$. 
    \item V sends back the one bit value of the inner product of $v_j$ and $r_j$ over GF[2], denoted as $\langle v_j , r_j \rangle$. 
    \item The server keeps a table T[x] of $2^l$ counters, corresponding to all possible $l$-bit values $x$ (initialized to zero on system setup).
    \item For every value of $x$ if $\langle x, r_j \rangle = \langle v_j , r_j \rangle$ the corresponding counter is incremented by one, otherwise it is decreased by one. Equality holds for exactly half of the values.
\end{enumerate}

Let the total number of service providers that ran the protocol be $N$ and $p$ be the frequency of the hash value $x$. The expected numbers of increments and decrements are $N(p + \frac{(1 - p)}{2})$ and $N\frac{(1 - p)}{2}$ respectively. The expected value of the counter is $E(T[x]) = pN$. For a threshold frequency $\tau$, the server publishes all $x$ values such as $T[x] > \tau N$. Each V checks if $H(\mathit{TID})$ is in the published hash values list. V can then act accordingly, either by discarding the certificate in the aggregation, or refusing entry to the user.

While this method can be used to detect Sybil-type B attacks there are a number of limitations to be considered. We consider three here: (1) To be practical we must efficiently increment and decrement all the counters with each certificate use, the naive approach to this requires a significant amount of memory and computational power. (2) A certificate hash will only be suspicious if, for $n$ the total number of times any certificate is used, that one hash is used $\Omega(\sqrt{n})$ times. This could allow an attacker to use each of many certificates slightly fewer than $\sqrt{n}$ times without (most of their certificates) being detected. (3) An adversarial service provider could submit false reports of certificates with the intention of causing false alarms, this could result in legitimate users having their certificates invalidated due to a hash collision with one of those certificates.

\end{document}